\documentclass[12pt]{article}  
\usepackage{amsmath}
\usepackage{url}
\usepackage[pdftex]{graphicx}
\usepackage[small,normal,bf,up]{caption2}

\usepackage[numbers]{natbib}

\title{Evolution of a Modular Software Network}
\author{Miguel A. Fortuna$^{1,2}$\footnote{To whom correspondence should be addressed. e-mail: fortuna@ebd.csic.es, Phone: +34 954466700, Fax: +34 954621125}, Juan A. Bonachela$^{2}$ and Simon A. Levin$^{2}$}

\date{ $^{1}$Integrative Ecology Group, Estaci\'on Biol\'ogica de Do\~nana, 
EBD-CSIC\\
Calle Am\'erico Vespucio s/n, 41092 Sevilla (Spain)\\ 
\vspace{0.2 in}
$^{2}$Department of Ecology and Evolutionary Biology\\
Princeton University, Princeton, NJ 08544-1003(USA)}

\begin{document}
\maketitle
\baselineskip=8.5 mm

\begin{center}
  {\sc Short Title:\\ Software Evolution}\\
  \vspace{0.25 in}
\end{center}

\vfill

\newpage
{\bf 
``Evolution behaves like a tinkerer'' (Fran\c{c}ois Jacob, 1977). Software systems provide a unique opportunity to understand biological processes using concepts from network theory. The Debian GNU/Linux operating system allows us to explore the evolution of a complex network in a novel way. The modular design detected during its growth is based on the reuse of existing code in order to minimize costs during programming. The increase of modularity experienced by the system over time has not counterbalanced the increase in incompatibilities between software packages within modules. This negative effect is far from being a failure of design. A random process of package installation shows that the higher the modularity the larger the fraction of packages working properly in a local computer. The decrease in the relative number of conflicts between packages from different modules avoids a failure in the functionality of one package spreading throughout the entire system. Some potential analogies with the evolutionary and ecological processes determining the structure of ecological networks of interacting species are discussed.
}

\newpage
Complex systems represented by networks pervade all sciences \cite{2}. Since the publication, ten years ago, of the first studies focused on the topological characterization and dynamical implications of networks of very different nature \cite{3,4,5,6,7,8,9}, little progress has been made on understanding the evolution of such complex systems (see however \cite{10,11,12}). Most studies assume that the architecture of these networks is static. However, on the World-Wide Web, pages and links are created and lost every minute \cite{13}. The structure of the current power grid depends on how it has grown over the years \cite{14}, and food webs are shaped continually through community assembly processes \cite{15}. Unraveling how these complex networks grow and change through time is a crucial task for understanding their long-term dynamics. 

Software systems, as computer operating systems, are under the constraints of hardware architecture and user requirements \cite{16}. Functionality is the main goal of software design. Developers need to make the system capable of accomplishing new tasks without excessive cost, so that modifying or adding a single feature does not require the update of preexisting code throughout the system. This ability to reuse existing code allows the system to build up in a modular and hierarchical fashion \cite{17}. The distributed and collaborative nature of software design, in which many individuals work only on small pieces of the whole system, requires developing a strategy to support software growth without losing functionality. Indeed, this attributed modular approach can enhance functionality \cite{18,19}.

Such design is expected to improve evolvability by limiting the interference between different functions. These interferences are a consequence of the software development process itself and may reduce the functional diversification of the operating system. For example, incompatibilities among functionally similar libraries required by different groups of programs may impede the correct installation of a complete set of software packages. Therefore, there is a trade-off between reusing many pieces of existing code and the emergence of incompatibilities among software packages. 

The Debian GNU/Linux operating system offers a unique opportunity to study the evolution of this trade-off over time, due to its package interaction system and its release schedule (see Fig.1). In Debian, most software packages reuse code of others in order to work properly (i.e. dependencies: package {\em i} needs package {\em j} for being functional) or have incompatibilities with other packages that impede the former to be installed in the same local computer (i.e. conflicts: package {\em i} prevents package {\em j} from being installed; see Methods).

In this paper we first characterize the evolving modular structure of the network of dependencies between software packages for the first ten releases of the Debian GNU/Linux operating system. Second, we explore the role of conflicts between packages in determining the functionality of the system by using a package installation process in a local computer. Last, we discuss potential parallelisms between the architecture and dynamics of software networks and that of ecological webs of interacting species.
 
\section*{Results}
We have compiled the binary i386 packages, including their dependencies and conflicts, of the first major stable versions of the Debian/GNU operating system released since the project began in 1993 (see Methods and Supplementary Information). The growth of the Debian/GNU Linux operating system from one release to the subsequent is summarized in three steps: some packages are deprecated, others are kept between versions, and new ones are added (see Fig.2). The number of packages that are deprecated between releases and those that persisted increased exponentially over time ($F_{1,7}=693.5, p<0.001$, and $F_{1,7}=165.2, p<0.001$, respectively, see Material and Methods). The number of new packages added in the most recent version was slightly smaller than in the previous one. If we discard it from the analysis, the number of new packages also increased exponentially over time ($F_{1,6}=216.9, p<0.001$). The total number of packages, dependencies and conflicts increased exponentially with each version, ranging from 448 to 28245 ($F_{1,8}=1117.8, p<0.001$), from 539 to 101521 ($F_{1,8}=603.1, p<0.001$), and from 28 to 4755 ($F_{1,8}=307.1, p<0.001$), respectively (see Table 1 in Supplementary Information). Data from the most recent release seem to indicate the beginning of an asymptotic stationary behavior for the growth of both, packages and interactions (their exclusion from the regression analysis increased the fit to $F_{1,7}=1064.6, p<0.001$, and to $F_{1,7}=466.8, p<0.001$, for dependencies and conflicts, respectively). Neither the ratio between the number of dependencies and the number of conflicts, nor the fraction of packages without any interactions showed a linear tendency over time ($F_{1,8}=1.5, p=0.263$; $21.2\pm5.18$ mean and standard deviation, and $F_{1,8}=0.078, p=0.787$; $0.132.\pm0.054$ mean and standard deviation, respectively).

The cumulative degree distribution for the outgoing dependencies (number of packages necessary for {\em i} to work) fit an exponential function ($F_{1,2}=114.2$, $F_{1,2}=358.1$, $F_{1,2}=1331.8$, $F_{1,3}=291.4$, $F_{1,3}=299.3$, $F_{1,4}=158.4$, $F_{1,5}=117.1$, $F_{1,5}=795.2$, $F_{1,5}=358.1$, and $F_{1,5}=280.8$, for all releases respectively; $p<0.001$ in all cases, see Fig.3). This means that there is a well-defined average number of packages that are used by others (see Fig.S1 in Supplementary Information). However, the cumulative degree distribution for the incoming dependencies (number of packages that need {\em i} to work) fit a power law function ($F_{1,5}=583.4$, $F_{1,5}=445.7$, $F_{1,7}=2403.3$, $F_{1,8}=3661.2$, $F_{1,9}=3278.6$, $F_{1,10}=945.6$, $F_{1,10}=1068.1$, $F_{1,1}=721.7$, and $F_{1,11}=900.4$, for all releases respectively; $p<0.001$ in all cases, see Fig.3). This means that a small number of packages are used by the vast majority while many programs are needed only by a few packages (see also Fig.S1 in Supplementary Information). In other words, the network of dependencies showed a scale-free distribution for the incoming dependencies over time, indicating that the new packages added on successive releases depended mainly on the most connected ones (i.e., those packages used by many others). 

The modular structure of the network of dependencies was statistically significant for all releases (ranging between $0.497$ and $0.564$; $p<0.001$ in all cases). The z-score obtained for allowing the comparison of the modularity across networks (see Methods) increased exponentially from the first version to the sixth ($9.664$, and $135.703$, respectively; $F_{1,4}=163.9$, $p<0.001$). Since then, it has remained around a lower stationary value ($44.555\pm11.669$, mean and standard deviation, respectively; $p=0.858$ for a linear regression). Although a significant linear relationship between the number of modules and the number of packages with dependencies is found for each version ($F_{1,8}=245.8, p<0.001$), the number of modules containing at least 5\% of the total number of packages for each version remained constant through time ($6.7\pm0.67$, mean and standard deviation, respectively; $F_{1,8}=1.5, p=0.250$; see Fig.S2 in Supplementary Information). Therefore, the new modules originated in subsequent releases contained few packages, indicating that the bulk of new packages added over time joined to the few large modules created in the earliest versions (see Fig.4 and Fig.S3 in Supplementary Information). 

Yet, the fraction of conflicts within modules increased linearly over time (ranging from $0.50$ to $0.74$; $F_{1,8}=30.45$, $p<0.001$) while the fraction of dependencies within modules remained constant ($0.677\pm0.031$, mean and standard deviation, respectively; $F_{1,8}=0.05$, $p=0.831$; see Fig.4 and Fig.S4 in Supplementary Information). Therefore, the increase in the modularity of the dependencies has not avoided the conflicts within modules during the exponential growth of the operating system. This means that, as the modular structure of the network of dependencies increased (up to the sixth release), the fraction of conflicts between modules decreased (Fig.4). Since then, although the modular structure has not grown significantly, the fraction of conflicts between modules has continued decreasing. 

The dynamical implications of this result are shown by a random process of package installation in a local computer (see Fig.1 and Methods). The fraction of packages that can be installed by a random process decreased linearly through time (ranging from $0.957$ to $0.711$; $F_{1,8}=40.1, p<0.001$). A priori, we might think that the higher the modularity of the network the lower the functionality of the operating system,  measured as number of packages installed from the pool of available software. However, other factors, such as the number of conflicts between packages (which also increased over time) may be responsible for the reduction in the fraction of software installed. To rule this effect out, we performed a random process of package installation in which the modular structure of the network of dependencies is deliberately broken by a local rewiring algorithm (see Methods). Hence, we can estimate the effect of the modular structure on the installation process. In almost all versions, the modularity of the network of dependencies increased significantly the fraction of packages installed in a local computer compared with what is expected from the randomization ($p<0.01$, except for releases 2.0 and 3.0, $p=0.14$ and $p=0.10$, respectively). The z-score calculated to compare this effect across releases (see Methods) did not show a significant linear increase over time until the release 3.0 ($1.685\pm0.569$, mean and standard deviation, respectively; $F_{1,5}=0.059$, $p=0.818$). Since then, the z-score increased notably (ranging from $17.9$ to $30.1$; see Fig.5). 

In summary, from the release 1.1 to the release 2.2 the significant exponential increase of the modularity of the network of dependencies (measured by the z-score) did not cause a significant positive effect on the fraction of software packages installed. However, from the release 3.1 to the last release analyzed (5.0), the lower and non-increasing modularity was responsible for a positive strong effect on the fraction of software installed in a local computer (Fig.5). Debian 3.1, released in 2005, proved to be a break point between these two opposite tendencies. Although versions 3.0 and 3.1 increased the amount of software to practically double the size of the previous release, Debian 3.1 updated 73\% of the packages found in Debian 3.0. These and other important changes are mainly the result of the long time elapsed since the previous version was released (almost 3 years, the longest interval between releases in the history of Debian; see http://en.wikipedia.org/wiki/Debian).
  
\section*{Discussion}
The increase of the modular structure of the operating system over time detected in this paper seems to be an effective strategy for allowing the growth of software minimizing the risk of collapse due to failures in the functionality of some packages. This strategy has also been reported for the ecological and evolutionary processes structuring food webs \cite{20}. The failure in the functionality of a software package, or the extinction of a species in an ecological community, would not propagate its negative effects to packages (species) from other modules, minimizing the risk of a collapse of the entire system. Therefore, understanding the evolution of a computer operating system can shed light on the evolutionary and ecological constraints shaping communities of interacting species. For example, we can investigate how species richness increases without jeopardizing the coexistence of the entire community. Minimizing the risks of competitive exclusion between species playing the same role in a community is equivalent to reducing software incompatibilities between modules of dependencies to increase functionally. The spatial segregation in the distribution of species represents an effective modular process analogous to the compartmentalization of the software network: it allows a higher regional species richness (software packages pool) at the expense of reducing local diversity minimizing competitive exclusion.

The Debian GNU/Linux operating system provides a unique opportunity to make this and other analogies within the evolutionary and ecological framework determining the structure of ecological networks of interacting species. Both processes occur at different time scales. On the evolutionary time scale, speciation and extinction, i.e. macroevolution, can be translated into the creation of new packages and the deprecation of those rendered obsolete from one version to the next. On the ecological time scale, colonization and local extinction, i.e. community assembly, would be equivalent to the package installation process in a local computer. Dependencies and conflicts between packages mimic predator-prey interactions and competitive exclusion relationships, respectively. Due to them, only a subset of the available packages can be installed in a computer, as only a subset of the species pool can coexist in a local ecological community. Moreover, there is an interplay between macroevolution and community assembly, because the interactions introduced by the new species (packages) alter the dynamics of the colonization/extinction (installation) in a local community (computer).

\section*{Conclusions}
During the exponential growth of the Debian GNU/Linux operating system, the reuse of existing code showed a scale-free distribution for the incoming dependencies and an exponential one for the outgoing dependencies. The modularity of the network of dependencies between packages as well and the number of structural modules increased over time. However, this increase in modularity did not avoid the increase in software incompatibilities within modules. Far for being a failure of software design, the modular structure of the network allows a larger fraction of the pool of available software to work properly in a local computer when the installation follows a random process. Decreasing conflicts between modules impedes the exclusion of entire modules of packages from the installation process. This positive effect of the modular structure was much larger in the three last releases, although the increase in modularity was not as high as it was for the first ones.  

Further research on network evolution and local assembly dynamics in this and in other engineer systems will open a new opportunity window for biologists and computer scientists to collaborate addressing fundamental problems in biology. Let us keep in mind the words of Uri Alon \cite{21}: ``The similarity between the creations of a tinkerer and engineer also raises a fundamental scientific challenge: understanding the laws of nature that unite evolved and designed systems''.

\section*{Methods}
\subsection*{Data set} 
In the Debian GNU/Linux operating system (www.debian.org) most software packages depend on or have conflicts with other packages in order to be installed on a local computer. By "dependencies" (package {\em i} depends on package {\em j}) we mean that package {\em j} has to be installed first on the computer for {\em i} to work. By "conflicts" (package {\em i} has a conflict with package {\em j}) we mean that package {\em i} cannot be installed if {\em j} is already on the computer. This does not necessarily mean that the package {\em j} has also a conflict with the package {\em i}: sometimes the package {\em j} is an improved version of the package {\em i} in a way that if {\em i} is already installed in the system then {\em j} improves it, but if {\em j} is installed then it already contains {\em i} and the latter cannot be installed.
We have compiled the list of software packages, along with the network of dependencies and conflicts, of the ten major versions released since 1996. The list of packages and interactions can be downloaded from the website of this journal (see Supplementary Information for more details). 

\subsection*{Statistical analysis}
We have performed exponential regressions to quantify the increase in the number of packages that were deprecated, the new ones that were added, and those that persisted among the ten releases analyzed. We also characterized the increase in the number of dependencies and conflicts through releases using exponential regressions. The change of the ratio between the number of dependencies and the number of conflicts through releases was tested using a linear regression. The fits of the cumulative degree distributions for dependencies and conflicts that are showed are those with the highest $F$-test statistic between the two applied functions (exponential and power law) using multiplicative bins (see Fig. S1). The increase of the fraction of dependencies and conflicts within modules through releases was tested using linear regressions. We used linear and exponential regressions to test the change in the z-score (obtained for allowing the comparison of the modularity across networks) through releases. Linear regressions were also used to characterize the relationship between the number of modules and the number of packages with dependencies through releases. Finally, the decrease in the number of packages installed by a random process through releases and its relationship with the z-score of the modularity were tested using linear regressions.

\subsection*{Modularity analysis} 
We have used a heuristic method, based on modularity optimization \cite{22}, to extract the modular structure of the network of dependencies of software packages constituting the different releases of the Debian GNU/Linux operating system. The "Louvain" method \cite{23} is a greedy algorithm implemented in C++ that allows one to study very large networks (the code is available at http://www.lambiotte.be). The excellent results in terms of modularity optimization given by the well-known "Netcarto" software based on simulated annealing \cite{24,25} is limited when dealing with large networks, where extracting modularity optimization is a computationally hard problem. It has been shown that the Louvain method outperforms Netcarto in terms of computation time \cite{23}. In addition, the Louvain method is able to reveal the potential hierarchical structure of the network, thereby giving access to different resolutions of community detection \cite{26}.
The statistical significance of the modularity was calculated by performing, for each release, 1000 randomizations of the network of dependencies keeping exactly the same number of dependencies per package, but reshuffling them randomly using a local rewiring algorithm \cite{27}). The p-value was calculated as the fraction of random networks with a modularity value equal to or higher than the value obtained for the compiled network. In order to rule out the differences (in terms of connectance, number of packages, etc.) in the comparison of the modularity across networks, we calculated a z-score defined as the modularity of the compiled network of dependencies minus the mean modularity of the randomizations, divided by the standard deviation of the randomizations. 
 
\subsection*{Local installation process} 
The aim of the local installation process is to calculate the distribution of the maximum number of packages that can be correctly installed in a computer by a random process of software installation.
We have performed 1000 replicates of the local installation process for each release of the Debian/GNU Linux operating system ensuring that the asymptotic behavior of the variance was reached. Only packages with interactions (dependencies and/or conflicts) have been used in the process, and no subset of basic packages has previously been installed (both conditions differ from the algorithm applied by Fortuna \& Meli\'an \cite{28}). The algorithm selects randomly a package and checks whether the packages it depends on or has a conflict with those that have already been installed. If the package has a conflict with an already installed one, it is discarded. If it has no conflict with installed packages, the algorithm checks whether any of the packages it depends on directly or indirectly (by successive dependencies), has been discarded or has a conflict with an already installed package. In that case, it is discarded too. Otherwise, it is installed with all the packages it depends directly as well as indirectly. The process continues until no more packages are available to be installed. In the few cases where a package depends on two packages having a reciprocal conflict (because one or the other is needed for the installation of the selected package), we choose randomly one of them and discard the other. 
The randomization of the network of dependencies used for testing the effect of the modularity on the local installation process was the same describe above (Modularity analysis). The number of conflicts between packages and the identity of who has a conflict with whom have remained unchanged, as in the compiled networks. We performed 1000 replicates of the installation process for each randomization, and generated 100 random networks of dependencies for each release. The fraction of random networks in which the fraction of packages installed was equal to or higher than the value obtained for the modular network was used as p-value. A z-score was calculated for comparing, across releases, the fraction of packages installed using the modular networks with that of randomizations. The z-score was defined as the mean fraction of packages installed using the modular network minus the mean fraction of packages installed using the randomizations, divided by the standard deviation of the randomizations.

\section*{Acknowledgments}
We thank Colin Twomey for useful discussions, and Nicholas Pippenger and Lu\'is A. Nunes Amaral for their comments and suggestions that have largely improved the ms. This work was funded by a Marie Curie International Outgoing Fellowship within the 7th European Community Framework Programme (to M.A.F.), and the Defense Advanced Research Projects Agency (DARPA) under grant HR0011-09-1-055 (to S.A.L.).

\newpage
\begin{figure*}[ht]
\begin{center}
\centerline{\includegraphics[bb=0 0 1024 768, width=0.7\textwidth]{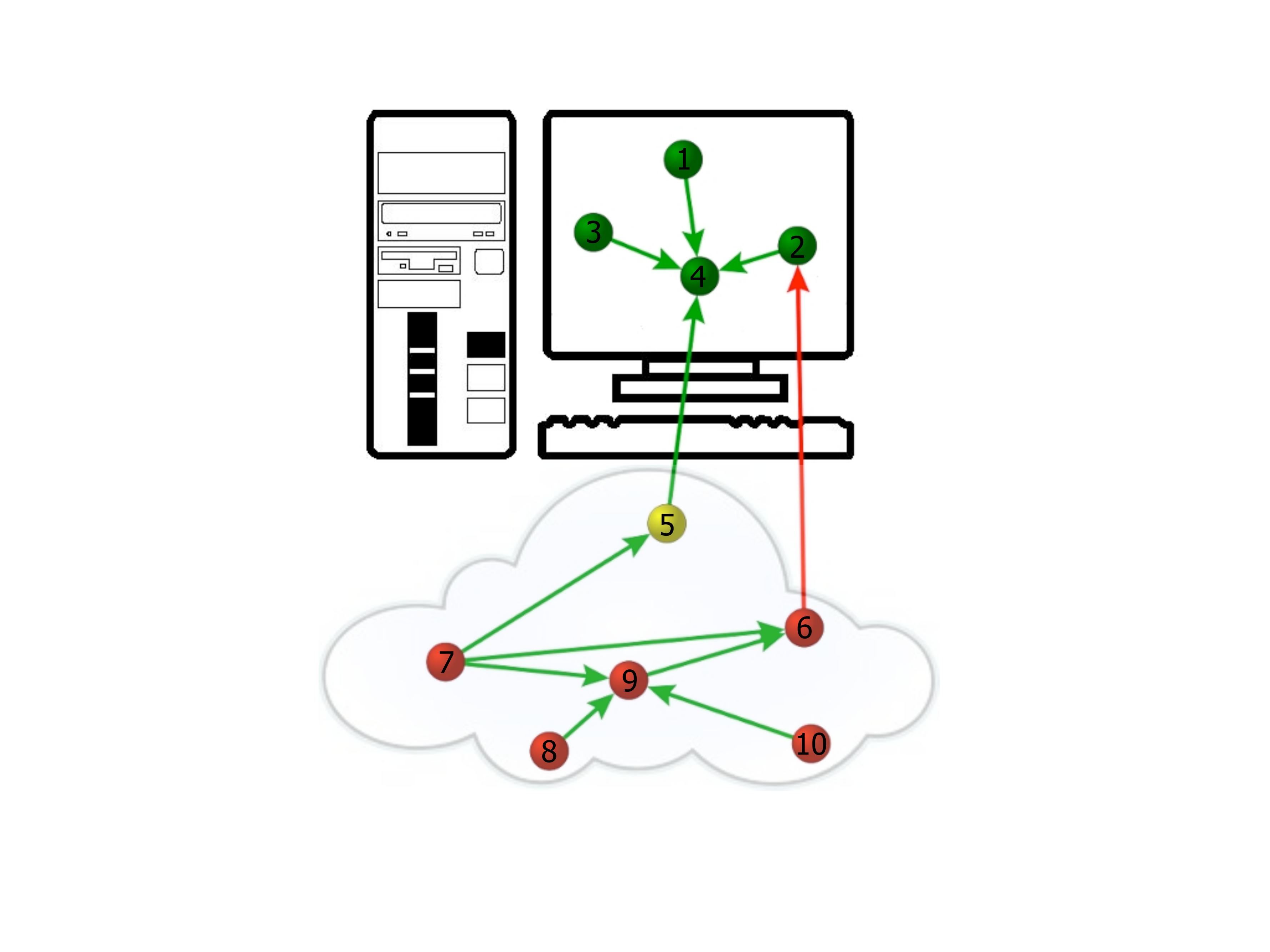}}
\caption{Dependencies and conflicts between packages during the installation of the Debian GNU/Linux operating system. Package {\em i} depends on package {\em j} (green arrows) if package {\em j} has to be installed first on the computer for {\em i} to work. Package {\em i} has a conflict with package {\em j} (red arrows) if package {\em i} cannot be installed if {\em j} is already on the computer. Packages, represented by nodes, are available for installation from the online servers or repositories (indicated in the figure by the cloud). The character of the interaction between packages determines which ones can be eventually installed on the computer. In this specific example, green nodes ({\em \#1-\#4}) represent packages already installed on the computer. For the network of packages in the cloud, only the package represented by the yellow node ({\em \#5}) can be installed on the computer. Package {\em \#6} has a conflict with an already installed package ({\em \#2}), and the remaining ones, ({\em \#7-\#10}), depend directly or indirectly on it. In this schematic local installation process, only half of the available packages can be installed on the computer. Different temporal sequences in the order of package installation will result in different sets of installed packages, or, in other words, functionalities of the operating system  (i.e. fraction of installed packages of the total number of available packages).}\label{Figure 1.}
\end{center}
\end{figure*}

\begin{figure*}[ht]
\begin{center}
\centerline{\includegraphics[bb=0 0 5243 948, width=0.9\textwidth]{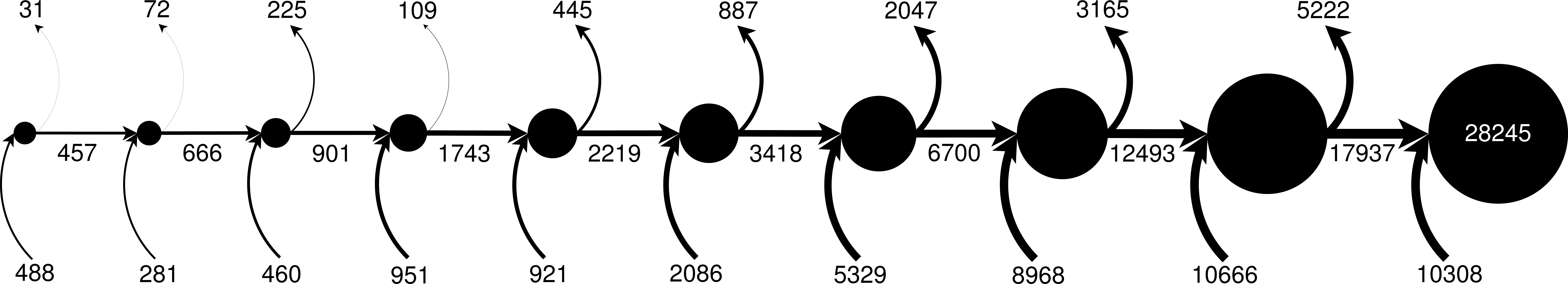}}
\caption{Schematic representation of the growth of the Debian GNU/Linux operating system through its first major releases. Circles depict releases and are arranged following the temporal sequence (from left to right). Their area is proportional to the logarithm of the number of packages in each release. Three arrows represent the transition between releases: the outgoing arrow indicates the number of packages that are deprecated from one release to the other; the incoming arrows represent the number of packages that give rise to the next release (some of them are updated from the previous release and the others are new packages). The number on the last node indicates the number of packages of the last analyzed release.}\label{Figure 2.}
\end{center}
\end{figure*}

\begin{figure*}[ht]
\begin{center}
\centerline{\includegraphics[bb=0 0 280 392, width=0.3\textwidth]{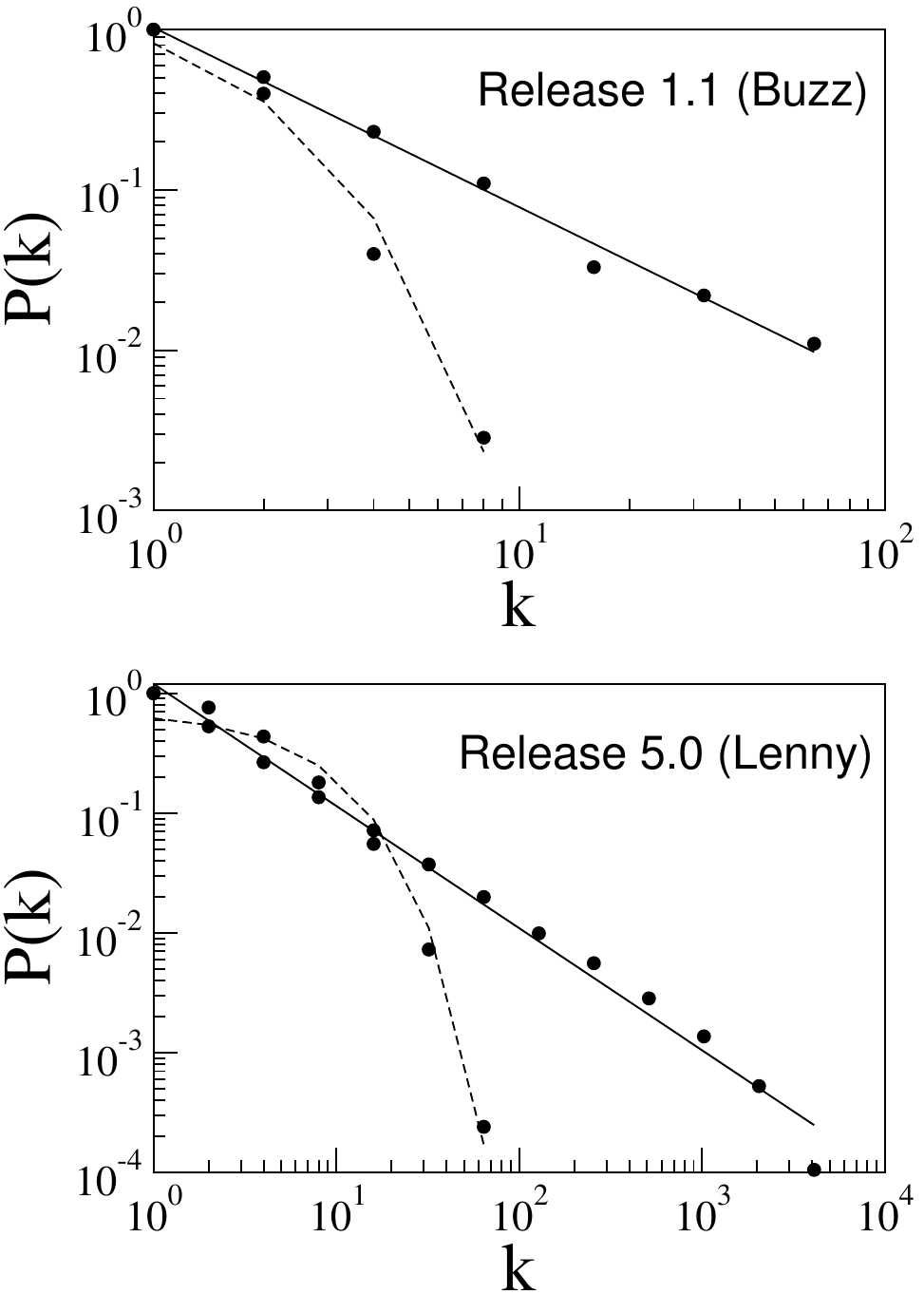}}
\caption{Cumulative degree distribution of the number of incoming (solid lines) and outgoing (dashed lines) dependencies for the software packages of the first and last releases (on top and on bottom, respectively) of the Debian GNU/Linux operating system analyzed here. The figures depict the probability, P($k$), for a package to depend on or to be needed by  at least, 1, 2, 3, ..., $k$ packages to work. Both axes are in logarithmic scale. In all cases the best fit for the outgoing dependencies is an exponential function while for the incoming dependencies is a power law (see also Fig.S1 in Supplementary Information).}\label{Figure 3.}
\end{center}
\end{figure*}

\begin{figure*}[ht]
\begin{center}
\centerline{\includegraphics[bb=0 0 1024 768, width=0.9\textwidth]{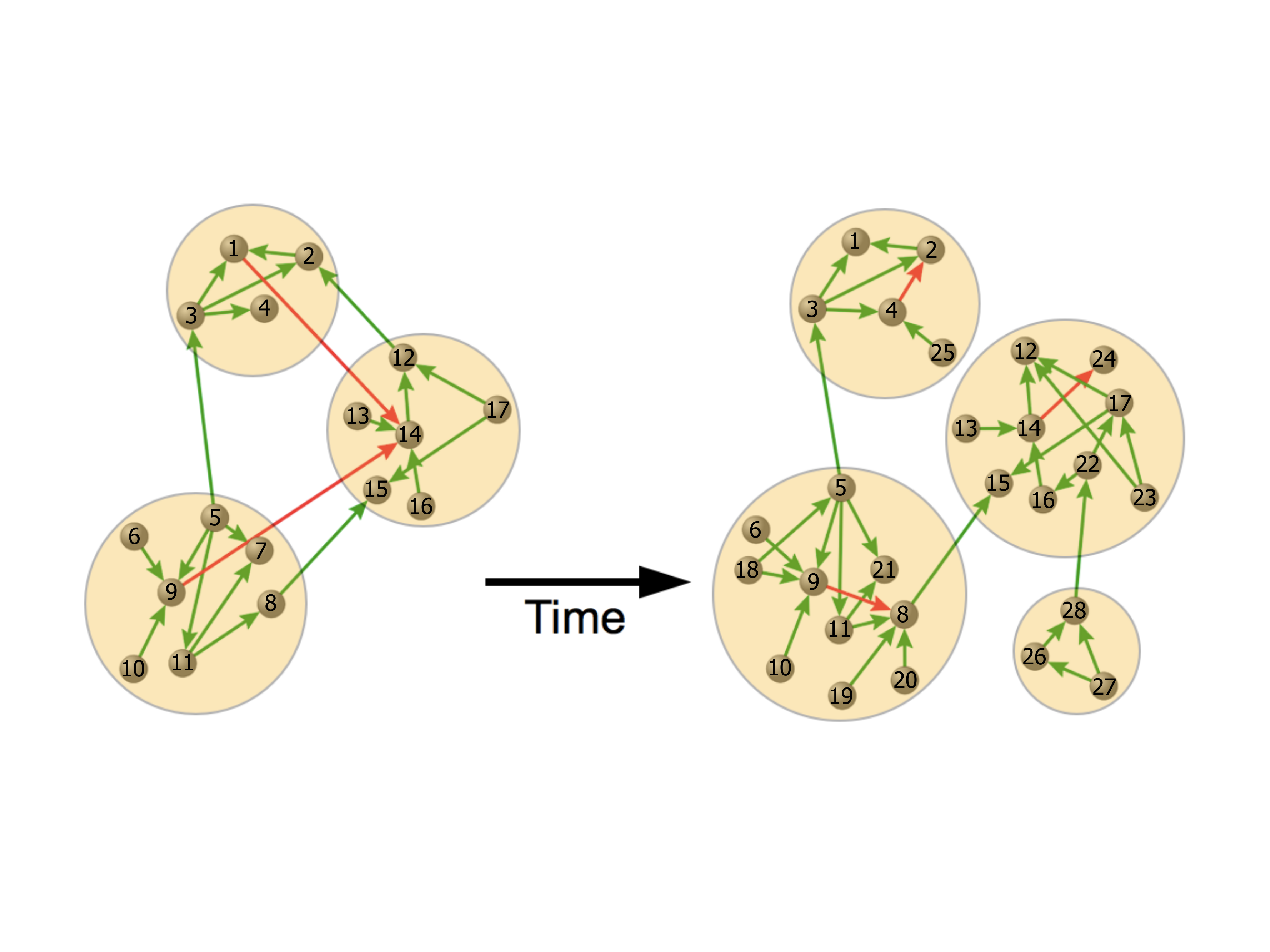}}
\caption{Evolution of the modular structure of the network of dependencies between packages of the Debian GNU/Linux operating system. Packages are represented by nodes. A green arrow from package {\em i} to package {\em j} indicates that package {\em i} depends on package {\em j}, and a red arrow indicates that package {\em i} has a conflict with package {\em j}. Packages within a module (depicted by a big circle) have many dependencies between themselves and only a few with packages from other modules. During the growth of the operating system the modular structure of the network of dependencies has increased: 1) the new packages added in successive releases depended mainly on previously existing packages within the same module, and hence, the size of the modules created in earlier releases increased over time; 2) the number of modules also increased, although the new modules consisted only of a few new packages; and 3) the relative number of dependencies between packages from different modules decreased. Moreover, the relative number of conflicts between packages from different modules decreased while those within modules increased through the different releases of the operating system.}\label{Figure 4.}
\end{center}
\end{figure*}

\begin{figure*}[ht]
\begin{center}
\centerline{\includegraphics[bb=0 0 339 453, width=0.3\textwidth]{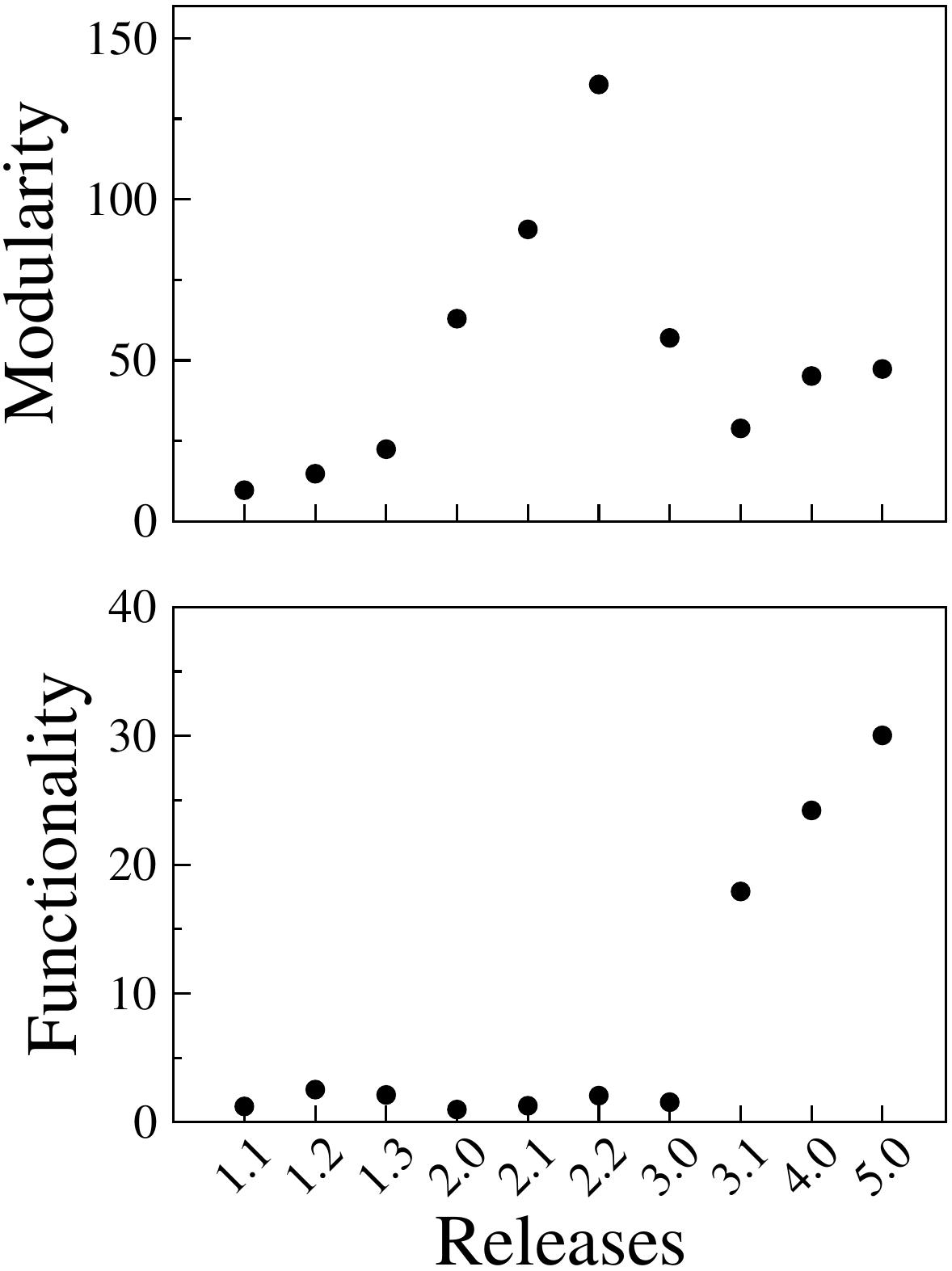}}
\caption{Changes of the modular structure (measured as the z-score of the modularity compared to a randomization of the modular structure) and functionality (measured as the z-score of the fraction of packages installed in a local computer compared to that installed from a randomization of the modular structure) for the network of dependencies of the releases of Debian GNU/Linux operating system analyzed here. The positive effects of the modular structure on the functionality shows up strongly in the last three releases (linear increase) although the exponential increase of the modularity happens in the first ones.}\label{Figure 5.}
\end{center}
\end{figure*}

\end{document}